



\def\Rrm{\hbox{\rm I\hskip -2pt R}}
\def\Crm{\hskip.1cm \hbox{\rm l\hskip -5.5pt C\/}}
\def\Zrm{\hbox{\bf  Z}}

\def\psaut{\vskip 5pt plus 1pt minus 1pt}
\def\saut{\vskip 10pt plus 2pt minus 3pt}
\def\gsaut{\vskip 20pt plus 3pt minus 4pt}

\font\sevenit=cmti7 \relax
\magnification=1200
 \vskip1cm
\centerline    {\bf CLOSEDNESS OF STAR PRODUCTS AND COHOMOLOGIES}
  \vskip1.5cm
 \centerline  {Mosh\'e FLATO and Daniel STERNHEIMER}
  \vskip.5cm
  \centerline {Physique Math\'ematique, Universit\'e de Bourgogne}
  \centerline {\sevenrm B.P. 138, F-21004 Dijon Cedex, France}
  \centerline {\sevenrm [e-mail: flato@u-bourgogne.fr
  \hskip.3cm daste@ccr.jussieu.fr]}
  \vskip.5cm
  \centerline{\it Dedicated to Bert Kostant with friendship and appreciation}
  \vskip.1cm
  \centerline {\sevenrm [ In: ``Lie Theory and Geometry: In Honor of
  B. Kostant", J.L. Brylinski et al. eds., pp. 241-259,
  Birkh\"auser, Boston, 1994]}
  \vskip.5cm
  \centerline {\bf Abstract}

  \sevenrm
  We first review the introduction of star products in connection with
  deformations of Poisson brackets and the various cohomologies that are
  related to them. Then we concentrate on what we have called ``closed
  star products" and their relations with cyclic cohomology and index
  theorems. Finally we shall explain how quantum groups, especially in
  their recent topological form, are in essence examples of star products.
  \tenrm
  \gsaut

\centerline{\bf 1. Introduction: Quantization}
\saut

\noindent {\bf 1.1 Geometry.} The setting of classical mechanics
in phase-space has long been a source of inspiration for mathematicians.
But (according to writing on a wall of the UCLA mathematics department
building) Goethe once said that ``Mathematicians are like Frenchmen:
they translate everything into their own language and henceforth
it is something completely different". Being French mathematicians,
we shall give here a flagrant illustration of that sentence, though not going
as far as Bert Kostant's cofounder of geometric quantization (Jean-Marie
Souriau) who derived symplectic formalism from the basic principles that
are the core of the French ``m\'ecanique rationnelle" established
by Lagrange.
\psaut
 The symplectic formalism is obvious in the Hamiltonian formulation on flat
 phase-space $\Rrm^{2\ell}$, and prompted in the fifties French
 mathematicians like Paulette Libermann and Georges Reeb to
 systematize the notion of symplectic manifold. Parallel to these
 developments  came the introduction of quantum  mechanics
 (first called ``m\'ecanique  ondulatoire"  in France under de Broglie's
 influence). And then (which brings  us  close to our subject here)
 Dirac [1] introduced (both in the classical  and in the quantum domain)
 his notion of ``constrained mechanics",  when external constraints restrict
 the degrees of freedom of phase-space.
 For mathematicians that is nothing but restricting phase-space to a
 submanifold of some $\Rrm^{2\ell}$ endowed with a Poisson manifold structure
 [2] (second class constraints give a symplectic submanifold);
 this restriction permits a nice and compact
 formulation of classical mechanics, but was of little help in the quantum
 case where people needed some reference to the canonical formalism on flat
 phase-space, as exemplified by the Weyl quantization procedure [3].
\psaut
Then quite naturally Kostant (coming from representation theory of Lie groups)
and  Souriau (coming from the symplectic formulation of classical mechanics)
introduced independently [4] what is now called {\it geometric quantization}.
The idea is to somehow select, via a polarization, a Lagrangian submanifold
$X$ of half the dimension of the symplectic manifold $W$ so that locally
 $W$ will look like $T^*X$, and quantization can be done on $L^2 (X)$;
 and in the meantime to work at the prequantization level on $L^2 (W)$.
 That idea, quite efficient for many group  representations,
ran into serious problems (now well-known) on the physical side;
in particular few observables were ``quantizable" in that sense.
\saut

\noindent{\bf{ 1.2 Deformations.}} The idea that the passage from one level
of physical theory to a more evolved one is done through the mathematical
notion of deformation became obvious only recently [5], but many
people certainly felt that something of this kind must occur.
On the space-time invariance level (Galileo to Poincar\'e to De Sitter)
it is simple to formulate because all objects are Lie groups. When
interactions (nonlinearities) occur, it is more intricate (and requires
the systematization of the notion of nonlinear representations [6]).
For quantum theories, in spite of hints in several expressions
(like ``classical limit"), the ``quantum jump" from functions
to operators remained.
Our approach, that started about 20 years ago [7-11] and is
now often referred to as {\it ``deformation quantization"}, showed that
there is an alternative (a priori more general) and autonomous formulation
in terms of ``star" products and brackets, deformations (in Gerstenhaber's
sense [12]) of the algebras of classical observables (Berezin [13] has
independently written a parallel formulation in the complex domain, but
not in terms of deformations).
\psaut
The importance of algebras of observables (especially $C^*$ algebras)
in quantum theories has even spilled into geometry with the
non-commutative geometry of Alain Connes [14] and its developments around
algebraic index theory (generalizing the Atiyah-Singer index theorems for
pseudo-differential operators), and with the exponential development
of quantum groups [15].
In this paper, after a survey of the origins (Section 2),
we shall (in Section 3) indicate that what we have called ``closed
star products" [16] permits a parallel treatment of the former, and
(in Section 4) that the latter, once formulated in a proper topological
vector space context, are essentially examples of star products.
In each case there are appropriate cohomologies to consider,
e.g., cyclic for closed star products and bialgebra cohomologies
for quantum groups, that are more specific than the traditional Hochschild
(and Chevalley) cohomologies. An alternative name for our approach could
thus be {\it ``cohomological quantization"}. It stresses the importance of
cohomology classes in all our approach and leads naturally to ideas
like ``cohomological renormalization" in field theory (when phase-space
is infinite-dimensional) where ``more finite"
cocycles can be obtained [17] by subtracting ``infinite" coboundaries
from cocycles to define star products equivalent to (but different from)
that of the normal ordering.
\gsaut

\centerline {\bf 2. Star Products and Cohomologies.}
\saut
{\bf 2.1  Deformations of Poisson Brackets.} Let $W$ be a
 symplectic manifold of dimension $2\ell$, with (closed)
 symplectic 2-form $\omega$; denote by $\Lambda$ the
 2-tensor dual to $\omega$ (the inner product by $-\omega$ defines
 an isomorphism $\mu$ between $TW$ and $T^*W$
 that extends to tensors). The Poisson bracket can be
 written as  $P(u,v) = i(\Lambda )(du \wedge dv) $
 for $u,v \in N = C^{\infty} (W) $. Some of the results
 described here are valid when $W$ is a Poisson manifold
 (where $\Lambda$ is given, with Schouten autobracket
 $[\Lambda , \Lambda] = 0$ --  the analogue of closedness for
 $\omega$ --  but not necessarily everywhere nonzero);
 the dimension need not be even and a number of results hold
 when the dimension is infinite, which is the case for field
 theory (Segal and Kostant [18] were among the first to consider
 seriously infinite-dimensional symplectic structures). We shall
 however not enter here into specifics of these questions.
 \psaut
 {\bf a.} A deformation of the Lie algebra $(N,P)$ is defined [12]
 by a formal series in a parameter $\nu \in \Crm $:
 $$[u,v]_\nu = P(u,v) +  \sum^\infty_{r=1}  {\nu}^r \hat{C}_r (u,v),
 \hskip.5cm {\rm for} \hskip.1cm u,v \in  N \hskip.1cm ({\rm or}
 \hskip.1cm N[[\nu]])  \eqno{(1)}$$
 such that the new bracket satisfies the Jacobi identity,
 where the $\hat{C}_r$ are 2-cochains (linear maps from $N \wedge N$ to $N$).
  In particular $\hat{C}_1$ must be a 2-cocycle for the
  Chevalley cohomology $\hat{H}^* (N,N)$ (for simplicity we
  shall write $\hat{H}^* (N)$, and similarly for Hochschild cohomology)
  of $(N,P)$.
  As usual, equivalences of deformations are classified at
  each step by $\hat{H}^2 (N)$ and the obstructions to extend
  a deformation from one step (in powers of $\nu$) to the
  next are given by $\hat{H}^3 (N)$ [11,12]. Whenever needed we shall
  put on the formal series space $ N[[ \nu ]] $ its natural
  $\nu$-adic topology.

  Moreover it can be shown that one gets consistent theories
  by restricting to differentiable (resp. 1-differentiable)
  cochains and cohomologies, when the $\hat{C}^r$ are
  restricted to being bidifferential operators (resp.
  bidifferential operators of order at most (1,1)).
  This has the advantage of giving finite-dimensional
  cohomologies with simple geometrical interpretation.
  In particular one has
  $$ {\rm dim}{\hat{H}}^2_{\rm {diff}} (N) = 1 +
  {\rm dim}{\hat{H}}^2_{\rm {1.diff}} (N)
  \hskip.5cm {\rm and} \hskip.5cm
   {\hat{H}}^p_{\rm {1.diff}} (N) =
  H^p (W, \Rrm ) \oplus H^{p-1} (W, \Rrm )  \eqno (2) $$
  if $\omega$  is exact (it can be smaller if not).

  One can also restrict to differentiable cochains that are null
  on constants (n.c. in short), i.e. $\hat{C}_r (u,v) = 0 $ whenever
  either $u$ or $v$ is constant, and again get a consistent
  theory. In that case $\hat{H}^2_{\rm {1.diff,nc}}(N) =
   H^2(W,\Rrm )$, the de Rham cohomology.
  Similar results hold for $\hat{H}^3_{\rm {diff}} (N)$,
  the obstructions space [19]; in particular
  $ \hat{H}^3_{\rm {diff,nc}} (N) $ is isomorphic to
  $H^1 (W) \oplus H^3 (W) $
  (for $\omega$ exact, modulo some condition on a 4-form; without it
  and/or without the n.c. condition the space may be slightly larger).
  This explains that, when the third Betti number of W,
  $b_3 = $ dim$H^3(W)$, vanishes, J. Vey was able to trace
  the obstructions inductively into the zero-class of $\hat{H}^3 (N)$
  and show the existence of such deformed brackets (the condition
  $b_3 = 0 $ is not necessary, as follows from the general
  existence theorems for star products that we quote later).
  Replacing in all the above ``differentiable" by ``local" gives
  essentially the same results for the cohomology [19].
  \psaut
  {\bf b.} In the differentiable n.c. case one thus gets that,
  if $b_2 = $dim$H^2(W) = 0$, there is (modulo equivalence) only
  one choice at each step, coming from the Chevalley cohomology
  class of the very special cocycle $S^3_{\Gamma}$ given,
  on any canonical chart $U$ of $W$, by

  $$ S^3_{\Gamma} (u,v) {\big\vert} _U = \Lambda^{i_1j_1} \Lambda^{i_2j_2}
  \Lambda^{i_3j_3} ({\cal L}(X_u)\Gamma)_{i_1i_2i_3}
  ({\cal L}(X_v)\Gamma)_{j_1j_2j_3}           \eqno (3) $$

  \noindent where ${\cal L}(X_u)$ is the Lie derivative in the direction
  of the Hamiltonian vector field $X_u = \mu^{-1}(du)$
  defined by $u \in N$ and $\Gamma$ is any symplectic connection
  ($\Gamma_{ijk}$ totally skew-symmetric; $i,j,k = 1,...,2{\ell}$)
  on $W$. The Chevalley cohomology class of $S^3_{\Gamma}$ is
  independent of the choice of $\Gamma$.
  On ${\Rrm}^{2{\ell}}$ (with the trivial flat connection)
  it coincides with $P^3$, when we denote by $P^r$ the $r^{\rm {th}}$
  power of the bidifferential operator $P$.
  It is [9-11] the pilot term for the Moyal bracket [21] M, given by (1)
  where $(2r+1)!{\hat C}_r = P^{2r+1}$, i.e. the sinh function of $P$
  (the only [11] function of $P$ giving a Lie algebra deformation).
  In the Weyl quantization procedure, M corresponds to the commutator
  of operators (when we take for deformation parameter
  $\nu = {1 \over 2}i \hbar ).$
  \gsaut

\noindent {\bf 2.2 Deformations of associative algebras.} On $N$
(or $N[[\nu]]$) we can consider the associative algebra defined by
the usual (pointwise) product of functions. Its deformations are governed
by the Hochschild cohomology $H^*(N)$, and here also it makes sense to
restrict to local or differentiable (n.c. or not) cochains and cohomologies.
All the latter cohomologies are in fact the same: $H^p(N) = {\wedge}^p (W)$,
the contravariant totally skew-symmetric $p$-tensors on $W$; if $b$ denotes
the Hochschild coboundary operator, any (local, etc.) $p$-cocycle $C$ is of
the form $C = D + bE$ with $D \in {\wedge}^p (W)$ and $E$ a (local, etc.)
$(p-1)$ cochain. This result was obtained in an algebraic context in [20].
\psaut
{\bf a.} In order to relate to the preceding theory and thereby reduce
 the (a priori huge) possibilities of choices, and also of course
 because this is the physically interesting case, we shall be interested
 only in deformations such that the corresponding commutator starts
 with the Poisson bracket $P$, what we call {\it ``star products"}:

 $$ u*v = uv + {\sum}^{\infty}_{r=1} {\nu}^r C_r(u,v), \hskip.5cm u,v \in N
 \hskip.1cm ({\rm or} \hskip.1cm N[[\nu]])                  \eqno (4)  $$
 $$ C_1(u,v) - C_1(v,u) = 2P(u,v)                  \eqno (5)  $$

 \noindent where the cochains $C_r$ are bilinear maps from $N \times N$ to $N$.
In the local (etc.) case, one necessarily has $C_1 = P + bT_1$, and
therefore any (local, etc.) star product is equivalent, via an equivalence
operator $T = I + {\nu}T_1$, $ T_1 $ a differential operator, to a star product
starting with $C_1 = P$. Note that, in the differentiable case, an equivalence
operator $T = I + {\sum}^{\infty}_{r=1} {\nu}^rT_r$ between two star products
is necessarily [11] given by a formal series of differential operators $T_r$
(n.c. in the n.c. case). Star products are always {\it nontrivial}
 deformations of the associative algebra $N$ because $P$ is a nontrivial
 2-cocycle for the Hochschild cohomology (a coboundary can never be a
 bidifferential operator of order (1,1)).

 The case when the cochains $C_r$ are differentiable and odd or even
 together with $r$ (what we call the {\it parity condition},
 $C_r(u,v) = (-1)^rC_r(v,u)$) is simpler and
we considered it first [11]. However the parity condition is not always needed,
and the differentiability condition is sometimes not completely satisfied. This
is especially the case when one deals with what we call ``star representations"
of (semi-simple) Lie groups $G$, by star products on coadjoint orbits.
There, the orbits being given by polynomial equations in the
vector space of the dual ${\cal G}^*$ of the Lie algebra ${\cal G}$ of $G$,
the $C_r$ will in general be bi-pseudodifferential
operators on the orbits. It would thus be of interest to introduce
another category of star products, when the cochains are algebraic functions
of bidifferential operators; the related cohomologies would probably
not be very different from the differentiable case ones.
On the other hand, restricting to 1-differentiable cochains is not of
much interest here (by opposition to the Lie algebra case [2])
since one then looses all connection to quantum theories
because the cochain $S^3_\Gamma$ is lost (the order of differentiation is
either 1 or unbounded).
\psaut
 {\bf b.} From (3) one gets a deformed bracket by taking the commutator
 ${1 \over {2\nu}}(u*v-v*u)$, which gives a Lie algebra deformation (1) with
 $2{\hat C}_{r-1}(u,v) = C_r(u,v) - C_r(v,u)$. In contradistinction with the
 Lie case, the Hochschild cohomology spaces are always huge
 but the choices for star products will be much more limited
 because of the associated Lie algebra deformations.

 In particular when the $C_r$ are differentiable and satisfy the parity
 condition, the $C_{2r+1}$ being n.c. (what is called a ``weak star
product"), any star product is equivalent [22] to a ``Vey star product",
one for which the $r!C_r$ have the same principal symbol as $P^r_{\Gamma}$,
the $r^{th}$ power of the Poisson bracket expressed with covariant derivatives
$\nabla$ relative to some symplectic connection $\Gamma$, i.e. on a local
chart $U$ (with summation convention on repeated indices):
$$ P^r_{\Gamma}(u,v) {\big\vert}_U = {\Lambda}^{i_1j_1}...{\Lambda}^{i_rj_r}
{\nabla}_{i_1...i_r}u {\nabla}_{j_1...j_r}v,  \hskip.3cm u,v \in N.
\eqno (6) $$
 For such products the most general form of the first terms are
   $$C_2 = P^2_{\Gamma} + bT_{(2)} \hskip.2cm {\rm and} \hskip.2cm
C_3 = S^3_{\Gamma} + {\Lambda}_2 + 3{\hat{b}}T_{(2)},  \eqno (7) $$
where $T_{(2)}$ is a differential operator of order at most 2,
$\hat{b}$ denotes the Chevalley coboundary operator and
${\Lambda}_2$ is a 2-tensor, image (under $\mu^{-1}$)
of a closed 2-form [19,22].
A somewhat general expression can also be given
for $C_4$ [19], but it is much more complicated. For higher
terms no explicit formula was published (Jacques Vey knew more
or less how to do it for $C_5$) but there exists an algorithmic
construction due to Fedosov [23] in terms of a symplectic connection that
gives a class of examples term by term.

What happens here (assuming the parity condition) is that the Lie
algebra (i.e. the odd cochains) determines inductively the star
product from which it originates; the only freedom is the possible
addition, at each even level, of multiples of $uv$ to the cochains
(and to the equivalence operators). The parity condition is of course
satisfied at level 0 and can (by equivalence) be assumed at level 1 for
star products, but the Lie algebra will in general (except when $b_2 = 0$)
give enough information only on the odd part of the cochains $C_r$.
\gsaut
\noindent {\bf 2.3. Existence, uniqueness and examples of star products.}

{\bf a. Existence.} On $\Rrm^{2\ell}$ with the flat symplectic connection
one has the Moyal star product and bracket:
$$ u*_Mv = {\rm exp}(\nu P)(u,v)  \hskip.5cm {\rm M}(u,v) =
{\nu}^{-1}{\rm sinh}(\nu P)(u,v).       \eqno (8)$$
 The idea is to take such star products $M_{\alpha}$ on Darboux charts
 $U_{\alpha}$  for any symplectic $W$ and glue them together. This cannot be
 done  brutally (when $b_3 \neq 0$ the topology of the manifold hits back).
 But $N[[\nu]]$ can be viewed as a space of flat sections in the bundle of
 formal Weyl algebras on the tangent spaces of $W$
 (a Weyl algebra is generated by the canonical commutation relations
 $ [{x^i} , x^{j} ] = 2{\nu}{\Lambda}^{ij}I $); a flat
 connection on that bundle is algorithmically constructed [23] starting
 from any symplectic connection on $W$.
 Pulling back the multiplication of sections gives a star product [24],
 which can also be seen as obtained by the juxtapositions
 of star products $T_\alpha M_\alpha$ on each
 $U_\alpha$, when the equivalence operators $T_\alpha$ are such that
 all $T_\alpha M_\alpha$ and $T_\beta M_\beta$ coincide on
 $U_\alpha \cap U_\beta$ (the Darboux covering is chosen locally finite).
 These star products can be taken to be differentiable n.c. (d.n.c. in short)
 and satisfying the parity condition.

 Earlier proofs of existence were done first in the case $b_3 = 0$, then
 for $W = T^*X$ with $X$ parallelizable, and shortly afterwards, for
 any symplectic (or regular Poisson) manifold; but that proof was
 essentially algebraic, while we now see better the underlying geometry.
 \psaut
 {\bf b. Uniqueness.} Formula (7) is very instructive about what happens.
  Indeed whenever we have two Lie algebra deformations equivalent to some
  order, after making them identical to that order by an equivalence,
  the difference of the cochains is a cocycle of the form
  given by $C_3$ in (7), {\it in the d.n.c. case} of course.
  Therefore we have at each step (for the bracket) at most
  $1+b_2$ choices modulo equivalence, and we see exactly where
  the second de Rham cohomology enters: the ``1" stands for
  the Moyal bracket, and the $b_2$-dimensional space comes
  from what we called in [7] ``inessential" 1-differentiable
  deformations that are obtained by deformations of the 2-tensor
  $\Lambda$, i.e., by deformations of the closed 2-form $\omega$
  (adding an exact 2-form gives an equivalent deformation).

  For star products ($P$ being a nontrivial Hochschild cocycle), the
  ``starting point" becomes the Moyal product, and the
  equivalence classes are classified by the second de Rham cohomology.
  Indeed we know now (cf. [23,24]) that it is always possible to
  avoid the obstructions; and if two star products are equivalent
  to order $k$, once they are made to coincide at that order the
  skew-symmetric part of their difference at order $k+1$ determines
  a closed 2-form that is exact iff they are equivalent to order $k+1$.
  (This follows from an argument due to S. Gutt, similar to those
  of [19,22]).

  In particular, also in the d.n.c. case and without the
  parity condition, {\it when $b_2 = 0$, the Moyal-Vey
  product is unique.}   In that case one
  can choose a star product satisfying the parity condition (denote
  its cochains by $C'_r$); any other d.n.c. star product (with cochains
  $C_r$) can then step by step be made equal to the chosen one by the
  abovementioned argument: at the first
  step where $C_k - C'_k$ is nonzero, it is of the form $D_k + bE_k$
  with $D_k$ a closed, thus exact, 2-form: $D_k = dF_k$ (here
  $D_k(u,v)$ is defined as $D_k(X_u \otimes X_v)$, and similarly
  $F_k(u) \equiv F_k(X_u)$); the equivalence will then be
  extended to the next order by $I - \nu^{k-1}F_k - \nu^kE_k$.
  \psaut
  {\bf c. Examples.} The various orderings considered in physics
   are the inverse image of the product (or commutator) of operators
   in $L^2(\Rrm^\ell)$ under the Weyl mappings
   $$ N \ni u \mapsto \Omega_w(u) = \int_{\Rrm^{2\ell}}
   \tilde{u}(\xi,\eta)
   {\rm exp}(i(P\xi + Q\eta)/\hbar)w(\xi,\eta)\omega^\ell
    \eqno (9) $$
  where $\tilde{u}$ is the inverse Fourier transform of $u$, $P$
  and $Q$ are operators satisfying the canonical commutation
  relations $[P_\alpha , Q_\alpha] = i\hbar\delta_{\alpha\beta}
   (\alpha, \beta = 1,...,\ell), w$ is a weight function,
   $2\nu = i\hbar$ and the integral is taken in the weak
   operator topology. Normal ordering corresponds to the weight
   $w(\xi,\eta) = {\rm exp}(-{1\over 4}(\xi^2 \pm \eta^2))$,
   standard ordering (the case of the usual pseudodifferential
   operators in mathematics) to $w(\xi,\eta) = {\rm exp}(-{i\over 2}\xi\eta)$
   and Weyl (symmetric) ordering to $w = 1$. Only the latter is such that
   $C_1 = P$ (e.g. standard ordering starts with the first half of $P$)
   but they are all mathematically equivalent via the Fourier transform of $w$.
   (Physically they give different spectra, when we define the star
   spectrum as indicated herebelow, for the image of
   most classical observables; in fact two isospectral star products
   are identical [25]).

   Other examples can be obtained from these products by various devices.
   For instance one can restrict to an open submanifold (like
   $T^*(\Rrm^\ell - \{0\})$), quotient it under the action of a group
   of symplectomorphisms and restrict to invariant functions; one can
   also transform by equivalences, or look (cf. below) for $G$-invariant
   star products. A variety of physical systems can thus be treated
   in an autonomous manner [11]. The simplest of course is the
   harmonic oscillator, which relates marvelously to the metaplectic
   group (dear to Bert Kostant). But other systems, such as the
   hydrogen atom, have also been treated from the beginning
   -- which is not the case of geometric quantization.

   An essential ingredient in physical applications is an autonomous
   spectral theory, with the spectrum defined as the support of
   the Fourier-Stieltjes transform of the star exponential
   ${\rm Exp}(tH) = \sum_{n=0}^\infty {1\over{n!}}(tH/i\hbar)^{*n}$,
   where the exponent $*n$ means the $n^{th}$ star power.
   Such a spectrum can even be defined in cases when operatorial
   quantization would give nonspectrable operators (e.g. symmetric
   with different deficiency indices). The notion of trace is also
   important here, and will bring us to closed star products.
   Interestingly enough, the trace of the star exponential of the
   harmonic oscillator was already obtained in 1960 by Julian
   Schwinger [26] (within conventional theory, of course).
   \psaut
 {\bf d. Groups.} $\Rrm^{2\ell}$ is the generic coadjoint orbit of the
 Heisenberg group in ${\cal H}^*_{\ell} =  {\Rrm}^{2\ell + 1}$;
 the uniqueness of Moyal parallels the uniqueness theorem of
 von Neumann, but this goes much further. For any Lie group $G$
 with Lie algebra ${\cal G}$ one has an autonomous notion of
 star representation. Every $x \in {\cal G}$
 can be considered as a function on ${\cal G}^*$
 and restricted to a function $u_x \in N(W)$ on a $G$-orbit
 (or a collection of orbits) $W$, so that $P(u_x,u_y)$ realizes
 the bracket $[x,y]_{\cal G}$. If we now take a star product on
 $W$ for which $[u_x,u_y]_\nu = P(u_x,u_y)$, what we call a
 {\it $G$-covariant star product}, the map $x \mapsto
  {1 \over 2}\nu^{-1}u_x$ will define a representation of the
 enveloping algebra ${\cal U(G)}$ in $N[\nu^{-1},\nu]]$, the
 space of formal series in $\nu$ and $\nu^{-1}$ (polynomial
 in the latter) with coefficients in $N$, endowed with the star
 product. This will give a {\it representation} of $G$ in
 $N[[\nu^{-1},\nu]]$ by the star exponential:
 $$ G \ni e^x \mapsto E(e^x) = {\rm Exp}(x) = \sum_{n=0}^\infty
 (n!)^{-1}(u_x/2\nu)^{*n}.     \eqno (10) $$

The star product is called {\it $G$-invariant} if $[u_x,v]_{\nu} = P(u_x,v)
\hskip.2cm \forall v \in N $, i.e. if the geometric action
of $G$ on $W$ defines an automorphism of the star product.
A whole theory of star representations has been developed
(see e.g. [27] for an early review). By now it includes an
autonomous development of nilpotent and solvable groups
(in a way adapted to the Plancherel formula), with a
correspondence (via star polarizations) with the usual
Kirillov and Kostant theories [28]; there the orbits are
(in the simply connected case) symplectomorphic to some
$\Rrm^{2\ell}$, and the Moyal product can be lifted to the orbits.
Star representations have also been obtained for compact groups and for
several series of representations of semi-simple Lie groups [29]
(including the holomorphic discrete series and some with unipotent orbits);
the cochains $C_r$ are here in general pseudodifferential.
Integration over $W$ of the star exponential (a kind of trace)
will give a scalar-valued distribution on $G$ that is nothing but
the character of the representation.
\gsaut
\centerline {\bf 3. Closed Star Products}
\psaut
\noindent {\bf 3.1 Trace and closed star products. Existence.}

{\bf a.} For Moyal product (weight $w = 1$ in (9)) one has, whenever
 $\Omega_1(u)$ is trace-class:
 $$ {\rm Tr}(\Omega_1(u)) = (2\pi\hbar)^{-\ell} \int u\omega^\ell
 \equiv {\cal T}_M(u)     \eqno (11) $$
 while for other orderings (like the standard ordering $S$) this
 formula is true only modulo higher powers of $\hbar$:
 ${\rm Tr}(\Omega_S(u)) = {\cal T}_M(u) + O(\hbar^{1-\ell})$.
 But for all of them the above-defined ${\cal T}_M$ has the
 property of a trace, i.e.,
 $$ {\cal T}(u*v) = {\cal T}(v*u).         \eqno (12) $$
 One even has (for the Moyal product, because of the skew-symmetry
 of the $\Lambda^{ij}$) that ${\cal T}_M(u*_Mv) = {\cal T}_M(uv)$.
 All these star products are what we call [16] {\it strongly closed}:
 $$ \int C_r(u,v) \omega^{\ell} = \int C_r(v,u) \omega^{\ell} \hskip1cm
 \forall r \hskip.1cm {\rm and} \hskip.1cm u,v \in N.     \eqno (13) $$

{\bf b.} The existence proofs of [24] can be made such that the
 star products  constructed are strongly closed.
 When $b_2 = 0$ the uniqueness (modulo equivalence) of star products
 shows that all d.n.c. star products are equivalent to a closed one
 (which exists). This is true on a general symplectic manifold:
 {\it All differentiable null on constants star products are, up
 to equivalence, strongly closed.}

 This is somewhat related to a recent result by O. Mathieu [30]
 that gives an often (not always, but always in degree 2)
 satisfied necessary and sufficient condition for the
existence of harmonic forms on compact symplectic manifolds, and thereby
counterexamples to a conjecture by J.L. Brylinski.

 To prove this result one considers the algebra
 $N[[\nu]]$ endowed with star product and restricts it (in order to
 get finite integrals) to $\cal D [[\nu]]$, where $\cal D$ denotes
 the $C^\infty$ functions with compact support on the manifold $W$.
 A trace on $\cal D [[\nu]]$ is then defined as a $\hskip.1cm
 \Crm [[\nu]]$-linear map $\cal T$ into $\hskip.1cm
 \Crm [\nu^{-1}, \nu]]$ satisfying (12). In the d.n.c. case
 it has been shown by Tsygan and Nest [31] that there exists (up to a factor)
 a unique trace on ${\cal D} [[\nu]]$.

 If one takes a locally finite covering of $W$ by Darboux charts $U_\alpha$
 (all the intersections of which are either empty or diffeomorphic to
 $\Rrm^{2\ell}$) and a partition of unity ($\rho_\alpha$) subordinate to it,
 this trace can be defined, in a consistent way [31], by
 $$ {\cal T}(u) = \sum_\alpha {\cal T}_\alpha (T_\alpha
 (\rho_\alpha * u))       \eqno (14)$$
 where ${\cal T}_\alpha$ is the Moyal trace (11) on the image of
 $U_\alpha$ in a standard $\Rrm^{2\ell}$ by coordinate maps, and
 $T_\alpha$ an equivalence that maps the given star product
 restricted to $U_\alpha$ into the Moyal product of the
 standard $\Rrm^{2\ell}$  (any self-equivalence of the Moyal
 product on $\Rrm^{2\ell}$ preserves the Moyal trace).
  It is thus of the form
  $$ {\cal T}(u) = (4i\pi\nu)^{-\ell} \int_W (Tu) \omega^\ell,
  \hskip.6cm {\rm with} \hskip.2cm
  T = I + \sum_{r=1}^\infty \nu^r T_r,      \eqno (15) $$
  the $T_r$ being differential operators, and this $T$ transforms
  the initial product into an equivalent one that is obviously closed.
  \psaut
   {\bf c.} In the last construction the d.n.c. assumption is not
   necessary, but relaxing it is not a trivial matter because it
   involves going beyond the differentiable case for the Hochschild
   cohomology and (for the n.c. assumption) because it is needed
   that 1 be a unity for the star algebra (differentiable star products
   can however be made d.n.c. by equivalence).  For more general
   star products the difference between general, closed and strongly
   closed star products should become manifest.

   A star product is said to be {\it closed} if (13) is supposed only
   for $r \leq \ell$, i.e. if the coefficient of $\nu^{\ell}$ in
   $(u*v - v*u)$ for all $u,v \in \cal D[[\nu]]$ has a vanishing integral.
  The reason for this definition is obvious from a glance at (11) if
  one remembers that in the algebraic index theorems of A. Connes [14],
  indices of operators are expressed as traces, and that star products
  permit an alternative and autonomous treatment of operator algebras
  directly on phase-space. Note that all star products on 2-dimensional
  manifolds are closed (because of (5)).
  \saut
  \noindent {\bf 3.2 Closed star products, cyclic cohomology
   and index theorems.}
   \psaut
   {\bf a. Cyclic cohomology} was introduced by A. Connes in connection
 with trace formulas for operators (the dual theory of cyclic
 homology was developed independently and in a different
 context by B. Tsygan) [32].
 The motivation is that cyclic cocycles are higher analogues
 of traces and thus make easier the computation of the index
 as the trace of some operator by giving it an algebraic setting.

 Let $\cal A$ be an algebra, $\cal A = \cal D [[\nu]]$ to fix ideas
 (but the notion of cyclic cohomology can be defined abstractly).
 To every $u \in \cal A$ one can associate ${\tilde u} \in {\cal A}^*$
 defined by
 $$ {\cal A}^* \ni {\tilde u}: {\cal A} \ni v \mapsto
 \int uv{\omega}^{\ell}.    \eqno (16) $$
 ${\cal A}$ acts on ${\cal A}^*$ by $(x{\phi}y)(v) = {\phi}(yvx)$, with
 $\phi \in {\cal A}^*$ and $v,x,y \in \cal A$. The map $u \mapsto
 {\tilde u}$ defines a map $C^p({\cal A}, {\cal A}) \rightarrow
 C^p({\cal A}, {\cal A}^*)$ compatible with the (Hochschild)
 coboundary operator $b$, and we can restrict to the space of cyclic
 cochains $C^p_{\lambda} \subset C^p ({\cal A}, {\cal A}^*)$, those
 that satisfy the cyclicity condition
 $$ {\tilde C}(u_1,...,u_p)(u_0) = (-1)^p{\tilde C}(u_2,...,u_p,u_0)(u_1).
     \eqno (17)$$
 The {\it cyclic cohomology} of $\cal A$, $HC^p(\cal A)$, is defined
 as the cohomology of the complex $(C^p_{\lambda}, b)$.
 Now, on the bicomplex $C^{n,m} = C^{n-m}(\cal A, {\cal A}^*)$
 for $n \geq m$ (defined as $\{ 0 \}$ for $n < m$), $b$ is of degree 1
 and one can define another operation $B$ of degree -1 that
 anticommutes with it $(bB = -Bb, B^2 = 0 = b^2)$. We refer to
 [14] and [16] for a precise definition in the general case; when
 $\gamma (1;u,v) = \int C(u,v) {\omega}^{\ell}$ is a normalized
 2-cochain $\gamma = {\tilde C} \in C^2 ({\cal A}, {\cal A}^*)$,
 $B$ can be defined by $B \gamma (u,v) = \gamma (1;u,v) - \gamma (1;v,u)$.
 This bicomplex permits to compute the cyclic cohomology (at each level
 $p$) by :
 $$ HC^p ({\cal A}) = ({\rm ker}b \cap {\rm ker}B) / b({\rm ker}B).
\eqno (18) $$  Obviously the closedness condition at order 2 of a star
product (4,5) is expressed by $B {\tilde C}_2 = 0 $. By standard deformation
theory [11] we know that the Hochschild 2-cocycle $C_1$ determines
a 3-cocycle $E_2$ that has to be of the form $bC_2$ (if the
 deformation extends to order 2) and therefore (if the star product
 is closed at order 2) ${\tilde E}_2 \in {\rm ker}b \cap {\rm ker}B$.
 Since modifying ${\tilde C}_2$ by an element of ker$B$ will not
 affect closedness, and since [11] the same story can be shifted
 step by step to any order, (18) shows us that the {\it obstructions
 to existence} of a $C_r$ yielding a star product closed at order
 $r \geq 2$ are given by $HC^3({\cal A} )$. Similarly, like in [11],
 the obstructions to extend an equivalence of closed star products
 from one step to the next are classified by $HC^2 ({\cal A})$.
 {\it Cyclic cohomology replaces Hochschild cohomology} for closed star
  products, and this will become especially important when the n.c.
  assumption is not satisfied.
  \psaut
 {\bf b. Character, and index theorems.} As in the Banach algebra
  case [33], which is a specification of the general framework
  developed by A. Connes [14], we can define here, when $*$ is closed:
  $$ {\varphi}_{2k}(u_0,...,u_{2k}) = \tau (u_0 * \theta (u_1,u_2) *...*
  \theta (u_{2k-1},u_{2k}))      \eqno (19) $$
  where $\ell \leq 2k \leq 2\ell$ (otherwise it is necessarily 0),
  $$ \theta (u_1,u_2) = u_1 * u_2 - u_1u_2 = \sum_{r=1}^{\infty}
  {\nu}^r C_r (u_1,u_2)      \eqno (20) $$
  is a quasi-homomorphism (that measures the noncommutativity of
  the $*$-algebra and is also a Hochschild 2-cocycle) and $\tau$
  is the trace defined by
  $$ \tau (u) = \int u_{\ell} \omega^{\ell},\hskip2cm
  u = \sum_{r=0}^{\infty} \nu^r u_r  \in  \cal D [[\nu]].  \eqno (21)$$
 This defines the {\it components of a cyclic cocycle} $\varphi$ in the $(b,B)$
 bicomplex on ${\cal D}$ that is called the {\it character} of the closed
 star product. In particular
  $$ \varphi_{2\ell} (u_0,...,u_{2 \ell}) =
  \int u_0 du_1 {\wedge}... {\wedge} du_{2 \ell}, \hskip1cm
  u_k \in {\cal D}.   \eqno (22) $$

 When $2 \ell$ = 4, a simple computation shows that the other
 component is $\varphi_2 = {\tilde C}_2$ and then $b\varphi_2 =
 -{1 \over 2}B\varphi_4$. But
 $HC^2({\cal D}) = Z_2 (W, \Crm) \oplus \Crm $
 (where $Z_2$ denotes the closed 2-dimensional currents on $W$).
 Therefore the {\it integrality condition} $<\varphi, K_0(\cal D)>
 \subset \Zrm$, necessary to have a deformation of $\cal D$ to
 the algebra of compact operators in a Hilbert space, what is
 traditionally called a quantization, {\it has no reason to be true}
 for a general closed star product.

 Now the pseudodifferential calculus on $W = T^*X$, with $X$
 compact Riemannian, gives [16] a closed star product, the character
 of which coincides with the character given by the trace on
 pseudodifferential operators, and therefore satisfies the
 integrality condition (up to a factor). The Atiyah-Singer index
 formula can then be recovered in an autonomous manner also in
 the star product formulation [14,16,23]. But the algebraic index
 formulas are valid in a much more general context [14,31,33].
 It is therefore natural to expect that the character of a general
 star product should make it possible to define a {\it continuous
 index}.

Recently a number of preprints have appeared (see e.g. [31] and [23])
deriving various proofs and generalizations of the Atiyah-Singer
 theorems using the (closed) star product formalism.

 \gsaut
 \centerline {\bf 4. Star Products and Quantum Groups}
 \saut
 \noindent {\bf 4.1 Topological Algebras.} The notion of quantum
 group has two dual aspects: the modification of commutation
 relations in Lie and enveloping algebras, and deformations of algebras
 of functions on a group (with star products). The latter gives by
 duality a coproduct deformation. In both cases {\it Hopf algebra}
 structures are considered, but there is a catch: except for
 finite-dimensional algebras, the algebraic dual of a Hopf
 algebra is not a Hopf algebra. The best way to circumvent this
 (largely overlooked) difficulty is to topologize the algebras in
 a proper way.
 \psaut
 {\bf a. Deformations revisited.} Let $A$ be a topological algebra.
  By this we mean an associative, Lie or Hopf algebra, or a
  bialgebra, endowed with a locally convex topology for which
  all needed algebraic laws are continuous. For simplicity we fix
  the base field to be the complex numbers $ \Crm$. Extending
  it to the ring $ \Crm [[\nu]]$ gives the module ${\tilde A} =
  A[[\nu]]$, on which we can consider various algebraic structures.

  A {\it deformation} of an algebra $A$ is a topologically free
  $ \Crm[[\nu]]$-algebra ${\tilde A}$ such that
  ${\tilde A}/\nu {\tilde A} \approx A$.
  For associative or Lie algebras this means that there exists
  a new product or bracket satisfying (4) or (1) (resp.). For a
  bialgebra (associative algebra $A$ with coproduct $\Delta :
  A \rightarrow A \otimes A$ and the obvious compatiblity relations),
  denoting by $\otimes_\nu$ the tensor product of $ \Crm [[\nu]]$-modules,
  one can identify ${\tilde A}{\hat{\otimes}}_\nu {\tilde A}$
  with $(A {\hat{\otimes}} A)[[\nu]]$, where ${\hat{\otimes}}$ denotes
  the algebraic tensor product completed with respect to some
  operator topology (projective for Fr\'echet nuclear topology
  e.g.), we similarly have a deformed coproduct
  $$ {\tilde \Delta } = \Delta + \sum_{r=1}^\infty \nu^r D_r,
  \hskip.2cm D_r \in {\cal L}(A, A {\hat{\otimes}}A)    \eqno (23)$$
  and here also an appropriate cohomology can be introduced [34-36].
  In the case of Hopf algebras, the deformed algebras will have the
  same unit and counit, but in general not the same antipode.
  As in the algebraic theory [12], {\it equivalence} of deformations
  has to be understood here as isomorphism of
  $\Crm [[\nu]]$-topological algebras (the isomorphism being the identity
  in degree 0 in $\nu$), and a deformation is said {\it trivial} if it is
  equivalent to that obtained by base field extension.
  \psaut
  {\bf b. The required objects.} In the beginning Kulish and Reshetikhin
  [37] discovered a strange modification of the ${\cal G} = s{\ell}(2)$
   Lie algebra, where the commutation relation of the two nilpotent
   generators is a sine in the semi-simple generator instead of
   being a multiple of it, which requires some completion of the
   enveloping algebra ${\cal U(G)}$. This was developed in the
   first half of the 80's by the Leningrad school of L. Faddeev,
   systematized by V. Drinfeld who developed the Hopf algebraic
   context and coined the extremely effective (though somewhat
   misleading) term of {\it quantum group} [15], and from the
   enveloping algebra point of view by Jimbo [38]. Shortly
   afterwards, Woronowicz [39] realized these models in the context
   of the noncommutative geometry of Alain Connes [14] by
   matrix pseudogroups, with coefficients in  $C^*$ algebras
   satisfying some relations.

  Let us take (for simplicity) a Poisson Lie group, a Lie group $G$
  with compatible Poisson structure i.e. a Poisson bracket $P$ on
  $N = C^{\infty}(G)$, considered as a bialgebra with coproduct
  defined by $\Delta u (g,g') = u(gg'), g,g' \in G,$ and satisfying
  $$ \Delta P (u,v) = P(\Delta u, \Delta v)  \hskip1cm {\rm where}
  \hskip.2cm u,v \in N.    \eqno (24)$$
  The enveloping algebra ${\cal U(G)}$ can be identified with
  distributions with (compact) support at the identity of $G$, thus is
  part of the topological dual $N'$ of $N$. But we shall need a space
  bigger than $N'$ for the quantized universal enveloping algebra
  ${\cal U_{\nu}(G})$, to include some infinite series in the Dirac $\delta$
  and its derivatives. Thus shall have to restrict to a subalgebra
  $H$ of $N$. When $G$ is compact we shall take the space $H$ of
  $G$-finite vectors for the regular representation.

  It is natural to look for topologies [40] such that both aspects
  will be in full duality, i.e. reflexive topologies. We also would
  like to avoid having too many problems with tensor product topologies
  that can be quite intricate; for instance, we need to identify
  $C^{\infty}(G \times G) = N \hat{\otimes} N$.

  When $G$ is a general Lie group, $N$ is Fr\'echet nuclear with dual
  ${\cal E}'$, the distributions with compact support, but ${\cal D}$
  (with dual ${\cal D}'$, all distributions) is only a (LF)-space, also
  nuclear. There is no simple candidate to replace the $G$-finite vectors
  of the compact case (the most likely are the analytic vectors for the
  regular, or a quasi-regular, representation). In the following we shall
  thus from now on restrict to the original setting of $G$ compact.
  \saut
  \noindent {\bf 4.2. Compact Topological Quantum Groups}\hskip.2cm [35].
  \psaut
  {\bf a. The Classical Objects.} For a compact Lie group $G$ we shall
  consider the following topological bialgebras (in fact, Hopf algebras):
  $$ H = \sum_{\rho \in \hat{G}}{\cal L}(V_{\rho}), \hskip.5cm
  {\rm and \hskip.1cm its \hskip.1cm dual}\hskip.1cm H' =
  \prod_{\rho \in \hat{G}} {\cal L}(V_{\rho})
   \supset {\cal D}'.    \eqno (25) $$
   Here $V_{\rho}$ is the isotypic component of type $\rho \in \hat{G}$
   in the Peter-Weyl decomposition of the (left or right) regular
   representation of $G$ in $L^2(G)$.
  $H$ is also called the space of coefficients, since it is the space
  of all coefficients of unitary irreducible representations (each
  isotypic representation being counted with its multiplicity,
  equal to its dimension). The enveloping algebra
  ${\cal U} = {\cal U(G)}$ is imbedded in $H'$ by
  $${\cal U} \ni x \mapsto i(x) = (\rho (x)) \in H'.    \eqno (26) $$
  This imbedding has a dense image in $H'$; the image is in fact
  in ${\cal D}'$ but is not dense for the ${\cal D}'$ topology.
  The product in ${\cal D}'$ is the convolution of distributions,
  the coproduct satisfies $\Delta (g) = g \otimes g$ for $g \in G$
  (identified with the Dirac $\delta$ at $g$), and the counit is
  the trivial representation. Since the objects will be the same
  in the ``deformed" case, only some composition laws being modified
  (exactly like in deformation quantization, of which it is in fact
  an example), we have discovered the initial group $G$ {\it ``hidden"}
  (like a hidden classical variable) in the compact quantum groups.

  All these algebras are what we call {\it well-behaved}: the
  underlying topological vector spaces are nuclear and either
  Fr\'echet or (DF). The importance of this notion comes from
  the fact that {\it the dual $A'$ of a well-behaved $A$ is
  well-behaved, and the bidual $A'' = A$.}
 \psaut
 {\bf b. The deformations.} First let us mention that duality
  and deformations work very well in the setting of well-behaved
  algebras: if ${\tilde A}$ is a deformation of $A$ well-behaved,
  its $\hskip.1cm \Crm[[\nu]]$-dual ${\tilde A}_{\nu}^*$ is a deformation
  of $A^*$ and two deformations are equivalent iff their duals
  are equivalent deformations.

  In view of the known models of quantum groups, we select a special
  type of bialgebra deformations, those we call [34,35] {\it preferred}:
  deformations of $N$ or $H$ with {\it unchanged coproduct}.
  Here this is not a real restriction because any coassociative
  deformation of $H$ or$N$ can (by equivalence) be made preferred (with a
  quasicocommutative and quasiassociative product).

  This follows by duality from the fact that for ${\cal D}'$ or $H'$,
  any associative algebra deformation is trivial, that these bialgebras
  are rigid (in the bialgebra category) and that any associative
  bialgebra deformation with unchanged product has coproduct
  $\tilde{\Delta}$ and antipode $\tilde{S}$ obtained from the undeformed
  structures by a similitude (expressing the ``quasi-" properties):
  $$ {\tilde{\Delta}} = {\tilde P} \Delta {\tilde P}^{-1},
  \hskip.3cm \tilde{S} = {\tilde a}S{\tilde a}^{-1}   \eqno (27)$$
  for some ${\tilde P} \in (A \hat{\otimes} A)[[\nu]]$ and
  ${\tilde a} \in A[[\nu]]$, with $A$ = ${\cal D}'$ or $H'$.
  When associative, the product is a {\it star product} that can
  (by equivalence [41]) be transformed into a (noninvariant)
  star product $*'$ satisfying
  $\Delta (u *' v) = \Delta u *' \Delta v, \hskip.1cm u,v \in N.$

  This general framework can be adapted to the various models. We
   refer to [35] for a thorough discussion. The Drinfeld and
   Faddeev-Reshetikhin-Takhtajan models fall exactly into this
   framework. An essential tool is the {\it Drinfeld isomorphisms}
   $\tilde{\varphi}$, (algebra) isomorphisms between a Hopf deformation
  ${\cal U}_{\nu}$ of $\cal U$ and ${\cal U}[[\nu]]$, which give (27).
  (Two Drinfeld isomorphisms give equivalent preferred deformations
  of $H$ that extend to preferred Hopf deformations of $N$).

  The Jimbo models [38] are somewhat aside because their classical
  limit is not ${\cal U}$ but ${\cal U(G)}$ extended by Rank($G$)
  parities, and this makes out of the deformed algebras nontrivial
  deformations.

  Finally we are now in position to have a good formulation
  of the {\it quantum double} [42]. If $A$ (resp. $A'$) denote $H'$
  or ${\cal D}'$ (resp. $H$ or $N$) or their deformed versions,
  the double is $D(A) = A' \bar{\otimes} A$; its dual is
  $D(A)' = A \hat{\otimes} A', D(A)'' = D(A)$ and all these
  algebras are rigid.
\psaut
{\bf c. Remark.} The above procedure can be adapted to noncompact
quantum groups. A first step in this direction can be found in [43].
\saut
\noindent {\bf Acknowledgements.} The authors want to thank A. Connes,
S. Gutt, P. Lecomte, G. Pinczon and B. Tsygan for worthy contributions
and A. Weinstein for useful comments. They also thank Ranee and Jean-Luc
Brylinski and Ann Kostant, without whom the conference would never have
been possible and this volume, hence this paper, would never have appeared.

 \gsaut
\centerline { REFERENCES}
\saut
\sevenrm
\noindent [1] P.A.M. Dirac, {\sevenit Lectures on Quantum Mechanics},
Belfer Graduate School of Sciences Monograph Series No. 2 (Yeshiva
University, New York, 1964).
\psaut
\noindent [2] M. Flato, A. Lichnerowicz and D. Sternheimer,
J. Math. Phys. {\sevenbf 17}, 1754-1762 (1976).
\psaut
\noindent [3] H. Weyl, {\sevenit The theory of groups and quantum mechanics},
Dover, New-York (1931).
\psaut
\noindent [4] B. Kostant, {\sevenit Quantization and unitary representations},
in: Lecture Notes in Math. {\sevenbf 170}, 87-208, Springer Verlag, Berlin
(1970).

\hskip -9pt  J.M. Souriau, {\sevenit Structure des syst\`emes dynamiques},
Dunod, Paris (1970).
\psaut
\noindent [5] M. Flato, Czech. J. Phys. {\sevenbf B32}, 472-475 (1982).
\psaut
\noindent [6] M. Flato, G. Pinczon and J. Simon, Ann. Sc. Ec. Norm. Sup.
${\hbox{4}}^e$ s\'erie, {\sevenbf 10}, 405-418 (1977).
\psaut
\noindent [7] M. Flato, A. Lichnerowicz and D. Sternheimer, C.R. Acad. Sc.
Paris {\sevenbf A 279}, 877-881 (1974). Compositio Mathematica {\sevenbf 31},
 47-82 (1975).
 \psaut
\noindent [8] M. Flato and D. Sternheimer, {\sevenit Deformations of Poisson
brackets, ...} in: {\sevenit Harmonic Analysis and Representations of
Semi-Simple Lie Groups}, J.A. Wolf et al. (eds.), 385-448, MPAM,
D. Reidel, Dordrecht (1980).
\psaut
\noindent [9] J. Vey, Comment. Math. Helv. {\sevenbf 50}, 421-454 (1975).
 \psaut
\noindent [10] M. Flato, A. Lichnerowicz and D. Sternheimer,
C.R. Acad. Sc. Paris, {\sevenbf A 283}, 19-24 (1976).
\psaut
\noindent [11] F. Bayen, M. Flato, C. Fronsdal, A. Lichnerowicz and
D. Sternheimer, Ann. Phys. (N.Y.), {\sevenbf 111}, 61-110 and 111-151 (1978).
\psaut
\noindent [12] M. Gerstenhaber, Ann. Math. {\sevenbf 79}, 59-90 (1964).
\psaut
\noindent [13] F. Berezin, Math. USSR Izv. {\sevenbf 8}, 1109-1165 (1974).
\psaut
\noindent [14] A. Connes. Publ. Math. IHES {\sevenbf 62}, 41-144 (1986);
{\sevenit G\'eom\'etrie Non-Commutative}, Inter\'editions, Paris (1990)
 [expanded English edition, preprint IHES/M/93/12, March 1993,
 to be published by Academic Press].
\psaut
\noindent [15] V.G. Drinfeld, {\sevenit Quantum Groups} in: {\sevenit Proc.
ICM86, Berkeley}, {\sevenbf 1}, 101-110, Amer. Math. Soc., Providence (1987).
\psaut
\noindent [16] A. Connes, M. Flato and D. Sternheimer, Lett. Math. Phys.
{\sevenbf 24}, 1-12 (1992).
 \psaut
\noindent [17] J. Dito, Lett. Math. Phys. {\sevenbf 27}, 73-80 (1993) and
{\sevenbf 20}, 125-134 (1992); and J. Math. Phys. {\sevenbf 33}, 791-801
 (1993).
 \psaut
\noindent [18] I.E. Segal, Symposia Mathematica, {\sevenbf 14}, 79-117
(1974).
               B. Kostant, ibid., 139-152.
 \psaut
\noindent [19] S. Gutt, {\sevenit D\'eformations formelles de l'alg\`ebre
des fonctions diff\'erentiables sur une vari\'et\'e symplectique},
Thesis, Bruxelles (1980); Lett. Math. Phys. {\sevenbf 3}, 297-309
(1979); Ann. Inst. Henri Poincar\'e {\sevenbf A 32},
1-31 (1980), and (with M. De Wilde and P. Lecomte) {\sevenbf A 40},
77-89 (1984).
 \psaut
\noindent [20] G. Hochschild, B. Kostant and A. Rosenberg, Trans. Am.
Math. Soc. {\sevenbf 102}, 383-406 (1962).
\psaut
\noindent [21] J.E. Moyal, Proc. Cambridge Phil. Soc.
{\sevenbf 45}, 99-124 (1949).

 A. Groenewold, Physica {\sevenbf 12}, 405-460 (1946).
 \psaut
\noindent [22] A. Lichnerowicz, Ann. Inst. Fourier Grenoble, {\sevenbf 32},
157-209 (1982).
 \psaut
\noindent [23] B. Fedosov, J. Diff. Geom. (in press) and pp. 129-136 in
{\sevenit ``Some topics of modern mathematics and their applications to
 problems of mathematical physics"} (in Russian, 1985). See also:
 Sov. Phys. Dokl. {\sevenbf 34}, 319-321 (1989); Advances in Partial
 Differential Equations (in press);
{\sevenit Proof of the Index Theorem for Deformation Quantization},
 (preprint Potsdam Max-Planck-Arbeitsgruppe, February 1994).

 C. Emmrich and A. Weinstein (Berkeley preprint, Nov. 1993);
 S. Gutt, Lectures at ICTP Workshop (Trieste, March 1993).
 \psaut
\noindent [24] H. Omori, Y. Maeda and A. Yoshioka, Lett. Math. Phys.
{\sevenbf 26}, 285-294 (1992); Adv. in Math. {\sevenbf 85}, 225-255 (1991).

 M. De Wilde and P. Lecomte, Note di Mat. {\sevenbf X} (1992); Lett.
 Math. Phys. {\sevenbf 7}, 487-496 (1983).
 \psaut
 \noindent [25] M. Cahen, M. Flato, S. Gutt and D. Sternheimer,
 J. Geom. Phys. {\sevenbf 2}, 35-48 (1985).
 \psaut
\noindent [26] J. Schwinger, Proc. Natl. Acad. Sci. {\sevenbf 46}, 1401-1415
(1960).
 \psaut
 \noindent [27] D. Sternheimer. Sem. Math. Sup. Montr\'eal {\sevenbf 102},
 260-293 (1986).
 \psaut
 \noindent [28] D. Arnal and J.C. Cortet, J. Geom. Phys. {\sevenbf 2},
 83-116 (1985); J. Funct. Anal. {\sevenbf 92}, 103-135 (1990).

 D. Arnal, J.C. Cortet and J. Ludwig, {\sevenit Moyal product and
 representations of solvable Lie groups} (to be published in J. Funct. Anal.).
 \psaut
 \noindent [29] D. Arnal, M. Cahen and S. Gutt, Bull. Soc. Math.
  Belg. {\sevenbf 41}, 207-227 (1989).
 \psaut
 \noindent [30] O. Mathieu, {\sevenit Harmonic cohomology classes of
 symplectic manifolds} (preprint, July 1993, to be published in
 Comment. Math. Helv.).
 \psaut
 \noindent [31] R. Nest and B. Tsygan, {\sevenit Algebraic Index Theorem}
 to appear in Commun. Math. Phys., and {\sevenit Algebraic Index Theorem
 For Families} to appear in Advances in Mathematics.
 B. Tsygan (private communication).
 \psaut
 \noindent [32] A. Connes, in: Math. Forschunginstitut Oberwolfach
 Tatungsbericht 41/81, Funktionalanalysis und C$^*$-Algebren,
 27-9/3-10 (1981).

 B. Tsygan, Russ. Math. Surveys {\sevenbf 38}, 198-199 (1983).
 \psaut
 \noindent [33] A.Connes, M. Gromov and H. Moscovici, C.R. Acad.
 Sci. Paris {\sevenbf I 310}, 273-277 (1990).

 A. Connes and H. Moscovici, Topology, {\sevenbf 20}, 345-388 (1990).
 \psaut
 \noindent [34] M. Gerstenhaber and S.D. Schack, Proc. Nat. Acad.
 Sci. USA, {\sevenbf 87}, 478-481 (1990); Contemp. Math. {\sevenbf 134},
 51-92 (1992).
 \psaut
 \noindent [35] P. Bonneau, M. Flato, M. Gerstenhaber and G. Pinczon,
 Commun. Math. Phys. {\sevenbf 161}, 125-156 (1994).
 \psaut
 \noindent [36] P. Bonneau, Lett. Math. Phys. {\sevenbf 26}, 277-280 (1992).
 \psaut
 \noindent [37] P.P. Kulish and N.Y. Reshetikhin, J. Soviet Math.
 {\sevenbf 23}, 24-35 (1983). (In Russian, Zap. Nauch. Sem. LOMI,
  {\sevenbf 101}, 101-110, 1981).
  \psaut
 \noindent [38] M. Jimbo, Lett. Math. Phys. {\sevenbf 10}, 63-69 (1985).
 \psaut
 \noindent [39] S.L. Woronowicz, Commun. Math. Phys. {\sevenbf 111},
 613-665 (1987).
 \psaut
 \noindent [40] F. Tr\`eves, {\sevenit Topological Vector Spaces,
 Distributions and Kernels}, Academic Press (1967).

 D. Sternheimer, {\sevenit Basic Notions in Topological Vector
 Spaces}, in: Proc. Adv. Summer Inst. in Math. Phys. (Istanbul
 1970, A. Barut ed.), 1-51, Studies in Math. Phys., D. Reidel,
 Dordrecht (1973).
 \psaut
 \noindent [41] L. Takhtajan, {\sevenit Introduction to quantum groups}
 in: Springer Lecture Notes in Physics, {\sevenbf 370}, 3-28 (1990).
 \psaut
\noindent [42] P. Bonneau, Reviews in Math. Phys. {\sevenbf 6}, 305-318 (1994).
\psaut
\noindent [43] F. Bidegain and G. Pinczon, Lett. Math. Phys. (to be published).
\end